\documentstyle[aps,multicol]{revtex}
\draft
\newcommand{\be}{\begin{equation}}
\newcommand{\ee}{\end{equation}}
\newcommand{\bea}{\begin{eqnarray}}
\newcommand{\eea}{\end{eqnarray}}

\def\cA{{\cal A}}

\def\Tr{{\rm Tr}\,}
\def\ep{\epsilon}
\def\bN{{\bar N}}

\def\bP{{\bar P}}
\def\bQ{{\bar Q}}
\def\bR{{\bar R}}

\newcommand{\px}{\! \not\!\partial}
\newcommand{\py}{\partial}

\begin{document}
  \begin{flushright} \begin{small}
     DTP-MSU/99-21,
     hep-th/9908133
  \end{small} \end{flushright}
\begin{center}
{\large \bf Vacuum Interpretation for Supergravity $M$--Branes}
\vskip.5cm
Chiang-Mei Chen,
Dmitri V. Gal'tsov
and Sergei A. Sharakin
\footnote{Email: chen@grg1.phys.msu.su, galtsov@grg.phys.msu.su,
sharakin@grg1.phys.msu.su}
\vskip.2cm
\smallskip
{\em Department of Theoretical Physics,
     Moscow State University, 119899 Moscow, Russia}
\vskip.1cm
\end{center}

\begin{abstract}
A non-local classical duality between the three-block truncated $11D$
supergravity and the $8D$ vacuum gravity with two commuting Killing
symmetries is established. The supergravity four-form field is generated
via an inverse dualisation of the corresponding Killing two-forms in six
dimensions. $11D$ supersymmetry condition is shown to be equivalent
to existence of covariantly constant spinors in eight dimensions. Thus any
solution to the vacuum Einstein equations in eight dimensions depending
on six coordinates and admitting Killing spinors have  supersymmetric
$11D$-supergravity counterparts. Using this duality we derive some new
brane solutions to $11D$-supergravity including $1/4$ supersymmetric
intersecting $M$-branes with a NUT parameter and a dyon solution joining
the $M2$ and $M5$--branes intersecting at a point.
\end{abstract}


\begin{multicols}{2}
\narrowtext

To reveal classical symmetries in dimensionally reduced supergravities
it is common to use dualisation of the antisymmetric forms
involved \cite{Ju81-98,St97}. The standard prescription consists in
using this procedure to maximally reduce the rank of forms in a given
dimension $D$, {\em i.e.} to dualise those forms whose rank $k$ is greater
than $D-k$.
Recently it was argued that deeper understanding of hidden symmetries
requires more general dualisation schemes, or even introduction
of double space of all forms both dualised and non-dualised with
the corresponding constraints \cite{CrJuLuPo98a}. Here we want
to explore a possibility provided by an {\em inverse dualisation}
($k<D-k$) to find a non-local correspondence between supergravities
involving higher rank antisymmetric forms and vacuum Einstein gravity
admitting isometries. The main idea is to dualise the Killing forms
associated with these isometries in order to generate higher rank
forms which could be then reinterpreted as matter fields of
suitable supergravity. For
definiteness we consider the correspondence between the three-block
truncation of $11D$-supergravity and $8D$ Einstein gravity with
two commuting isometries.

Consider the bosonic sector of $11D$ supergravity
\bea
S_{11} &=& \int d^{11}x \sqrt{-\hat g_{11}} \left\{ \hat R_{11}
   - \frac1{48} \hat F^2_{[4]} \right\} \nonumber\\
  &-& \frac16 \int \hat F_{[4]} \wedge \hat F_{[4]} \wedge\hat A_{[3]},
      \label{11d}
\eea
assuming for the space--times the following three--block structure
$M^{11} = M^2(z^a) \times M^3(y^i) \times M^{1,5}(x^\mu)$, where
$M^2$ and $M^3$ are conformally flat euclidean spaces, $M^{1,5}$ is the
pseudoriemannian six-dimensional space--time and it is understood that
all variables depend only on coordinates $x^\mu$:
\bea
ds_{11}^2 &=& g_2^\frac12 \delta_{ab} dz^a dz^b
   + g_3^\frac13 \delta_{ij}  dy^i dy^j \nonumber \\
  &+& g_2^{-\frac14} g_3^{-\frac14} g_{\mu\nu} dx^\mu dx^\nu.
   \label{MA}
\eea
Here the scalars $g_2, g_3$ and the six-dimensional metric $g_{\mu\nu}$
are functions of $x^\mu$. The corresponding consistent truncation of $11D$
supergravity involves the following ansatz for the $11D$ four--form
$\hat F_{[4]} = d \hat A_{[3]}$:
\bea
&&\hat F_{[4]\mu\nu ab} = \epsilon_{ab} F_{\mu\nu},\quad
\hat F_{[4]\mu ijk} = \epsilon_{ijk} \partial_\mu \kappa, \nonumber\\
&&\hat F_{[4]\mu\nu\lambda\tau} = H_{\mu\nu\lambda\tau},
   \label{FA}
\eea
where $H = d B$ and $F = d A$ are the {\em six--dimensional} four--form and
two--forms respectively.

It is straightforward to show that the equations
of motion for the six--dimensional variables can be derived from the
following action
\bea
S_6 &=& \int \, \sqrt{-g_6} \; d^6x \Bigl\{ R_6
   - \frac12 e^{2\phi} (\partial \kappa)^2
   - \frac12 (\partial \phi)^2
   - \frac13 (\partial \psi)^2 \nonumber \\
  &-& \frac14 e^{-\phi} \left(  e^\psi F^2
   + \frac1{12} e^{-\psi} H^2 \right) \Bigr\}
   + \kappa \int F \wedge H,
     \label{EA}
\eea
where
\be
\phi = -\frac12 \ln g_3, \quad
\psi = -\frac34 \ln g_2 - \frac14 \ln g_3. \label{Phi}
\ee

Our main result consists in the following

{\bf Theorem:}
{\em Any solution of $8D$ vacuum Einstein gravity with two commuting
spacelike isometries gives rise to several solutions of the $11D$
supergravity of the form (\ref{MA},\ref{FA}).
If, in addition, the $8D$ space--time
admits covariantly constant spinors, the corresponding $11D$ solutions
have unbroken supersymmetries.}

To prove these assertions, first we note that the scalar fields
$\kappa, \phi$ parameterize the coset $SL(2,R)/SO(2)$. To put the action
in a manifestly $SL(2,R)$--covariant form one has to dualise the four--form
$H$ as follows:
\be
H_{\mu\nu\lambda\tau} = \frac12 \sqrt{-g} e^{\psi+\phi}
  \epsilon_{\mu\nu\lambda\tau\rho\sigma}
  \left( *H^{\rho\sigma} + \kappa F^{\rho\sigma} \right).
\ee
The equations of motion in the new variables can be derived from the action
\bea
S_6 &=& \int \, \Big\{ R_6 - \frac13 (\partial \psi)^2
  - \frac14 e^\psi {\cal F}^T_{\mu\nu} {\cal M} {\cal F}^{\mu\nu} \nonumber\\
 &+& \frac14 \Tr (\partial{\cal M}\partial{\cal M}^{-1} ) \Big\}\,
 \sqrt{-g_6}\; d^6x,
 \label{EAM}
\eea
where
\be
{\cal M} = \left( \begin{array}{cc}
                  e^{-\phi}+\kappa^2 e^\phi & \kappa e^\phi \\
                  \kappa e^\phi & e^\phi
                  \end{array} \right), \qquad
{\cal F_{\mu\nu}} = \left( \begin{array}{c}
                  F_{\mu\nu} \\ {} *H_{\mu\nu}
                  \end{array} \right).
\ee
This action has a manifest $R \times SL(2,R)$ symmetry:
\bea \label{sl}
& {\cal M}' = {\cal G}^T {\cal M} {\cal G}, \quad
  {\cal F}_{\mu\nu}' = {\cal G}^{-1} {\cal F}_{\mu\nu}, \quad
& {\cal G} \in SL(2,R), \\
& \psi' = \psi + t, \quad
  {\cal F}_{\mu\nu}' = e^{-\frac{t}2} {\cal F}_{\mu\nu}, \quad
& t \in R,
\eea
which can be used to generate new supergravity solutions.

Now let us start with the $8D$ Einstein theory on space--times
with two spacelike Killing symmetries, in which case the $8D$ metric
can be presented as
\bea \label{d6}
ds_8^2 &=& h_{ab} (d\zeta^a+\cA_{\mu}^a dx^\mu)(d\zeta^b+\cA_{\nu}^b dx^\nu)
    \nonumber \\
    &+& (\det h)^{-\frac14} g_{\mu\nu} dx^\mu dx^\nu,
\eea
where $h_{ab}$, $2 \times 2$ matrix, and $\cA_{\mu}^a$ are two Kaluza-Klein
(KK) vectors depending on $x^\mu$ only. Under the KK reduction to six
dimensions one gets two KK two--forms, and three scalar moduli parameterizing
the metric $h_{ab}$. The eight-dimensional Einstein equations in terms
of the six-dimensional variables may be cast exactly into the form (\ref{EAM})
provided the following identification holds:
\bea
h_{ab} &=& e^{\frac23\psi} \left( \begin{array}{cc}
                  e^{-\phi}+\kappa^2 e^\phi & \kappa e^\phi \\
                  \kappa e^\phi & e^\phi
                  \end{array} \right), \nonumber \\
F &=&d \cA^1 , \quad *H= d \cA^2.\label{DT}
\eea
This completes the proof of the bosonic part of the theorem.
Note that the four--form $H_{\mu\nu\lambda\tau}$
of the $11D$ supergravity is generated from the KK two--form
by {\em inverse dualisation}. Possible permutation of two
Killing vectors in (\ref{d6}) will lead to different $11D$ solutions.
Moreover, if the actual number of commuting spacelike isometries of the $8D$
solution is greater than two, one can generate several different $11D$
solutions via different choice of the ordered pair of  $\zeta^a$--directions
in (\ref{d6}).

Consider now the $11D$ supergravity condition for vanishing of the
supersymmetry variation of the gravitino field (Killing spinor equation) in
a bosonic background satisfying the ansatz (\ref{MA},\ref{FA}):
\be
D_M\ep_{11} \!+\! \frac1{288}({\Gamma_M}^{\bN\bP\bQ\bR}
   \!-\! 8\delta_M^{\bar N}\Gamma^{\bP\bQ\bR})
   \hat F_{\bN\bP\bQ\bR}\ep_{11} \!=\! 0. \label{SE}
\ee
Here $\ep_{11}$ is a 32-component Majorana spinor, and the orthonormal
components of the gamma-matrices and the four--form are denoted by bar.
Consistently with the three--block decomposition for the metric one
assumes the $2\times 3\times 6$ split for the spinor
$\ep_{11} = \ep_2 \times \ep_3 \times \ep_6$ and the $11D$ gamma--matrices
\bea
\Gamma_{\bar a} &=& \rho_{\bar a} \times \sigma_{\bar 0}
    \times \gamma_{\bar 7}, \quad
  \Gamma_{\bar i}=\rho_{\bar z}\times\sigma_{\bar i}\times\gamma_{\bar 7},
    \nonumber\\
\Gamma_{\bar\alpha} &=& \rho_{\bar 0} \times \sigma_{\bar 0}
    \times \gamma_{\bar\alpha},
\eea
where $\rho_{\bar a}$, $\sigma_{\bar i}$ are two sets of $2\times 2$ Pauli
matrices, $a=x,y$, $i=x,y,z$, $\rho_{\bar 0}$ and $\sigma_{\bar 0}$
are unit matrices and $\gamma_{\bar\alpha}$ are $6D$ gamma--matrices:
$\{\gamma_{\bar\alpha},\gamma_{\bar\beta}\}=2\eta_{\bar\alpha\bar\beta}$.
The chiral operator is defined as
$\gamma_{\bar 7}=\gamma_{\bar 0}\gamma_{\bar 1}
\gamma_{\bar 2}\gamma_{\bar 3}\gamma_{\bar 4}\gamma_{\bar 5}$,
so that $\gamma_{\bar 7}^2=1$ and
$\{\gamma_{\bar 7},\gamma_{\bar\alpha}\}=0$.

Without loss of generality  one can  assume that $\ep_2$
and $\ep_3$  are constant  spinors.  Then  after reduction
the Eq.(\ref{SE}) is cast into a set of three
equations one of which is the $6D$ differential equation for the spinor
$\ep_6$:
\bea
\Bigl\{ 4D_{\bar\alpha} &-& \frac13 \py_{\bar\alpha} (\phi+2\psi)
   - \frac13 \gamma_{\bar\alpha}{}^{\bar\beta} \py_{\bar\beta} \psi
   + e^{\frac{\psi+\phi}2} \gamma_{\bar7} (*H_{\bar\alpha}
  \!+\! \kappa F_{\bar\alpha}) \nonumber\\
  &-& i s_z \left( e^{\phi} \gamma_{\bar 7} \py_{\bar\alpha} \kappa
   + e^{\frac{\psi-\phi}2} F_{\bar\alpha} \right) \Bigr\} \ep_6 = 0,
    \label{s6}
\eea
where $D_{\bar\alpha}$ is the six-dimensional spinor covariant derivative
and the constant spinor $\ep_2$ is split into the
chiral projections $\rho_{\bar z}\ep_2=s_z\ep_2$,
while two others are the algebraic equations for $\ep_6$ which
impose consistency conditions on the bosonic background:
\bea
\Bigl\{ \px (\phi-\frac23 \psi) - i s_z \left( e^{\phi}
     \gamma_{\bar 7} \px \kappa
   + e^{\frac{\psi-\phi}2}{\bf F} \right) \Bigr\} \ep_6 \!&=&\! 0,
     \label{s4} \\
\Bigl\{ \px (\phi\!+\!\frac23 \psi) \!+\! e^{\frac{\psi+\phi}2}
     \gamma_{\bar 7} (*{\bf H} \!+\! \kappa {\bf F})
  \!-\! i s_z e^{\phi} \gamma_{\bar 7} \px \kappa \Bigr\} \ep_6
  \!&=&\! 0. \label{s5}
\eea
Here  the following notation is used
\bea
\px &=& \gamma^{\bar\alpha}\partial_{\bar\alpha}
    = \gamma^{\bar\alpha}e^{\mu}_{\bar\alpha} \partial_{\mu}, \nonumber\\
{\bf F} &=& \frac12 F_{\bar\alpha\bar\beta} \gamma^{\bar\alpha\bar\beta},
    \quad
F_{\bar\alpha} = \frac12[\gamma_{\bar\alpha}, {\bf F}]
    = F_{\bar\alpha\bar\beta} \gamma^{\bar\beta},
\eea
and similarly for $*H$. In deriving the Eqs. (\ref{s6}-\ref{s5}) all terms
containing the products of three, four and five gamma matrices were
eliminated via the Fierz identities.

One can show that these equations are equivalent to the geometric Killing
spinor equation in $D=8$:
\be\label{s8}
D_M \ep_8 = \py_M \ep_8 + \frac14 \Omega^{\bar A\bar B}{}_M
    {\cal Y}_{\bar A\bar B} \ep_8 = 0,
\ee
where $\ep_8$ is 16-component spinor, the symbol
${\cal Y}_{\bar A\bar B}$ denotes an antisymmetric product
of the $8D$ gamma--matrices and $D_{M}$ stands for the spinor
covariant derivative with the riemannian connection corresponding to the
metric (\ref{d6}). Splitting the $8D$ spinor as $\ep_8=\ep'_2\times\ep'_6$,
where $\ep'_2$ is a constant spinor, and decomposing the gamma-matrices
${\cal Y}_{\bar a}=\tau_{\bar a}\times\gamma_{\bar 7}, \;
{\cal Y}_{\bar\alpha}=\tau_{\bar 0}\times\gamma_{\bar\alpha}$,
($\tau_{\bar a}$ is the Pauli matrix) one
derives from (\ref{s8}) a six-dimensional spinor equation ($M=\mu$) and some
matrix constraint equation ($M=a$). The subsequent derivation (details will
be given elsewhere) involves an unitary transformation of the corresponding
matrix finally arriving at the Eqs.(\ref{s4}-\ref{s5}), and establishing
the following correspondence between six-dimensional spinors:
\be
\ep'_6 = e^{-\frac{\phi+2\psi}{12}} \ep_6.
\ee
This proves the fermionic part of our theorem. Due to the
integrability condition for the Eq.(\ref{s8}), the $8D$ metrics
admitting Killing spinors will be Ricci-flat. Also, it is worth noting that
the  supersymmetry conditions are invariant under symmetry transformations
(\ref{sl}).

We do not intend to explore here the general classes of $11D$ solutions
which can be generated via this duality, but rather proceed with
some physically meaningful examples.
The simplest examples are the non--rotating black
$M2$--brane and $M5$--brane. Consider the $8D$ vacuum metric
\begin{eqnarray}
ds_8^2 &=& H \left( d\zeta_1+A_t dt \right)^2 + d\zeta_2^2
   - H^{-1} f dt^2 \nonumber\\
  &+& dx_1^2 + \cdots + dx_k^2 + f^{-1} dr^2 + r^2 d\Omega_{4-k}, \label{8d}
\end{eqnarray}
where $k$ in an integer, $(0 \le k \le 2)$ and
\begin{equation}
H = 1 \!+\! \frac{2\delta}{r^{3-k}}, \;
f = 1 \!-\! \frac{2m}{r^{3-k}}, \;
A_t = \frac{2\sqrt{\delta(m+\delta)}}{r^{3-k}H}.
\end{equation}
Two commuting Killing vector fields may be chosen either as
$(\partial_{\zeta_1}, \partial_{\zeta_2})$ or as
$(\partial_{\zeta_2}, \partial_{\zeta_1})$.
In the first case, performing
the steps described above, we arrive at the $M2$-brane solution
\begin{eqnarray}
ds_{M2}^2 &=&
  H^{-\frac23}\left( -\! f dt^2 \!+\! dz_1^2 \!+\! dz_2^2 \right)
   + H^{\frac13} \Bigl( dy_1^2 \!+\! dy_2^2 \!+\! dy_3^2 \nonumber\\
  &+& dx_1^2 + \cdots + dx_k^2
   + f^{-1} dr^2 + r^2 d\Omega_{4-k} \Bigr), \label{BM2}
\end{eqnarray}
and $\hat A_{tz_1z_2} = A_t$.
In the second case one gets the $M5$-brane
\begin{eqnarray}
ds_{M5}^2 &=&
  H^{-\frac13} \left( -f dt^2 + dz_1^2 + dz_2^2 + dy_1^2
   + dy_2^2 + dy_3^2 \right) \nonumber\\
  &+& H^{\frac23}\left( dx_1^2 \!+\! \cdots \!+\! dx_k^2
  \!+\! f^{-1} dr^2 \!+\! r^2 d\Omega_{4-k} \right),
  \label{BM5}
\end{eqnarray}
with the standard three--form field \cite{Co96a}.
For $m=0$ the $8D$ metric (\ref{8d}) admits Killing spinors
$\ep_8=H^{-\frac14} \ep_0$ (in cartesian local frame), where $\ep_0$ is
a constant spinor satifying ${\cal Y}_{\bar y\bar t}\ep_0=\ep_0$.
This clearly corresponds to the BPS saturated $M$-branes.
In a similar way one can obtain rotating branes, solutions endowed
with waves and KK monopoles as well as some their intersections.

The above construction illustrates that our mapping from $8D$ to $11D$
is one to many, {\em i.e.} an unique $8D$ vacuum solution can have
several $11D$ supergravity counterparts depending on the choice and
the order of two Killing vectors in eight dimensions.
This provides a simple way to proliferate $11D$ solutions: taking
one known solution one can find its $8D$ vacuum partner and then go back
to eleven dimensions with a different choice of the Killing vectors involved.
For example, two well-known five--dimensional KK
solutions, the Brinkmann wave and the KK-monopole, can be combined in a
six--dimensional vacuum solution which has six $11D$ counterparts including
all pairwise intersections of four basic solutions: $M2$--brane,
$M5$--brane, wave and monopole except the $M2 \bot M5$--brane.
One can also find the NUT--generalisations of the rotating $M$-branes in a
way similar to that used in the five-dimensional theory \cite{ChGaMaSh99}.

New interesting solution looking like a non-standard intersection of the
$M2$ and $M5$ branes at a point (not over a string as demanded by the
usual intersection rules)
can be obtained starting with the $5D$ Kaluza--Klein dyon.
The most general dyonic black hole solution of $5D$ KK
theory was found by Gibbons and Wiltshire \cite{GiWi86} and its
rotating version by Cl\'ement \cite{Cl86a} and Rasheed \cite{Ra95}.
In the static case this is a three--parameter family (with mass, electric and
magnetic charges $m, q, p$):
\bea
ds_8^2 &=& \frac{B}{A} \left( d\zeta_1 + {\cal A}_\mu dx^\mu \right)^2
   + d\zeta_2^2 + dx_1^2 + dx_2^2 \nonumber \\
  &+& \sqrt{\frac{A}{B}} \left\{ \frac{-\Delta}{\sqrt{AB}} dt^2
   + \sqrt{AB} \left( \frac{dr^2}{\Delta} + d\Omega_2 \right) \right\},
\eea
where $A=r_-^2-2dp^2/d_-,\; B=r_+^2-2dq^2/d_+,\; r_{\pm}=r\pm d/\sqrt{3},\;
d_{\pm}=d\pm \sqrt{3}m,\;\Delta = r^2 - 2 m r + p^2 + q^2 - d^2$, and
the KK vector field is given by ${\cal A}_t = {C}/{B}, \;
{\cal A}_\phi = 2p\cos\theta,\;C = 2 qr_-$.
The scalar charge $d$ satisfies the cubic equation
\be \label{cu}
q^2/d_+ + p^2/d_- = 2 d / 3.
\ee

Using the relations (\ref{DT}, \ref{Phi}) and then (\ref{MA},\ref{FA}),
one obtains the $11D$ solution
\bea
ds_{11}^2 &=& \left( \frac{B}{A} \right)^{-\frac16} \Bigl\{
    \frac{-\Delta}{\sqrt{AB}} dt^2
  + \sqrt{AB} \left( \frac{dr^2}{\Delta} + d\Omega_2 \right) \Bigr\}
    \nonumber \\
 &+& \left( \frac{B}{A} \right)^{-\frac23}
     \left( dz_1^2 + dz_2^2 \right) \nonumber\\
 &+& \left( \frac{B}{A} \right)^\frac13 \left( dy_1^2 + dy_2^2
  + dy_3^2 + dx_1^2 + dx_2^2 \right), \nonumber \\
\hat A_{tz_1z_2} &=& {\cal A}_t, \qquad
\hat A_{\phi z_1z_2} = {\cal A}_\phi. \label{M2uM5}
\eea

This solution for $p=0$  reduces to the usual $M2$--brane while for
$q=0$ --- to the $M5$--brane
(localized only on a part of the full transverse space). However, for both
$q,\,p$ non-zero, this is a new dyonic brane: contrary to the Costa
$M2 \subset M5$--brane \cite{Co96a}, for which the electric brane lies
totally within the magnetic one, and contrary to the intersecting
$M2 \bot M5$--brane (with a common string), now the common part
of the world-volumes is zero-dimensional (we suggest the name
$M2 \cup M5$). The electric/magnetic charge densities are $Q=8\pi q$
and  $P=8\pi p$ respectively.

Let us derive the supersymmetry conditions for this solution.
As $\kappa = *H = 0$ for it and the Eq.(\ref{s5})
is satisfied identically, the only constraint on the functions
$A,B,C,\Delta$ is due to the Eq.(\ref{s4}) which may be represented as
\be  \label{sus}
\left\{ I_6 + i s_z \left[ \alpha(r) \gamma^{\bar t}
   - \beta(r) \gamma^{\bar r\bar\theta\bar\varphi} \right]
     \right\} \ep_6 = 0,
\ee
where
$$
\alpha(r) = \frac{B^2\py_r(C/B)}{A^{3/2}\Delta^{1/2}\py_r(B/A)}, \;\;
\beta(r) = \frac{2pB^{3/2}}{A^2\Delta^{1/2}\py_r(B/A)}.
$$
Non-zero solution to (\ref{sus}) can exist if only $\alpha^2(r)+\beta^2(r)=1$,
what implies asymptotically
\be
p^2 + q^2 = 4d^2/3.
\ee
Combining this with the cubic constraint (\ref{cu}) we find that either
i) $p=0$, $m=d/\sqrt{3}$, $q=2d/\sqrt{3}$, corresponding to an extremal
$M2$-brane (\ref{BM2}), or
ii) $q=0$, $m=-d/\sqrt{3}$, $p=2d/\sqrt{3}$, an extremal $M5$-brane
(\ref{BM5}), otherwise the constraints lead to pure vacuum.
So the only nontrivial supersymmetric members of the
family of composite $M2\cup M5$-branes are the single extremal $M2$ and
$M5$-branes.

One can also obtain the rotating version of $M2\cup M5$--brane
(with one rotational parameter) through the following transformation:
\bea
dt &\to& dt + \omega d\phi, \quad
  \Delta \to \Delta + a^2,  \nonumber \\
A &\to& A + a^2\cos^2\theta
  + \frac{2 j p q \cos\theta}{d_+^2/3 - q^2},
  \nonumber \\
B &\to& B + a^2\cos^2\theta
  - \frac{2 j p q \cos\theta}{d_-^2/3 - p^2},
  \nonumber \\
C &\to& C - \frac{2 j p (d_+/\sqrt3) \cos\theta}
                 {d_-^2/3 - p^2},
\eea
where
$$
\omega = \frac{2j\sin^2\theta}{\Delta\!-\!a^2\sin^2\theta} \Bigl[ r\!-\!m
 \!+\! \frac{d_+(m^2\!+\!d^2\!-\!p^2\!-\!q^2)}
           {\sqrt{3}(d_+^2/3-q^2)} \Bigr].
$$
In terms of new variables one component of the KK vector potential,
$\cA_t$, remains the same  while another becomes
\bea
\cA_\phi &=& \frac{C}{B}\omega
  + \frac{2p\Delta\cos\theta}{\Delta-a^2\sin^2\theta} \nonumber \\
 &+& \frac{2jq\sin^2\theta \left[ (r d_- - md)/\sqrt{3}
           + p^2 + q^2 - d^2 \right]}
         {(\Delta-a^2\sin^2\theta)\left( d_+^2/3-q^2 \right)}.
\eea
The angular momentum of the system, $j$, is given by
\be
j^2 = a^2 \frac{\left( d_+^2/3-q^2 \right)\left( d_-^2/3-p^2 \right)}
               {m^2 + d^2 - p^2 - q^2}.
\ee

The drawback of our generating technique is that the $11D$ branes obtained
can not {\em localize} on all coordinates of transverse space, since
our basic ansatz (\ref{MA}) is not general enough. However one still can use
it to explore possible geometric $11D$ structures and then to try to localize
solutions in a straightforward way. Here we present without
derivation (details will be given in a forthcoming paper) a new partially
localized intersecting $M2$ branes with a NUT parameter:
\bea
d \hat s_{11}^2
  \!&=&\! (F_1 F_2)^\frac13 \Bigl\{\!-\!(F_1 F_2)^{-1}
      (dt\!-\!\omega_i d x^i)^2 \!+\! F_1^{-1} (dz_1^2 \!+\! dz_2^2)
      \nonumber\\
  \!&+&\! F_2^{-1} (dw_1^2 \!+\! dw_2^2)
   + dy_1^2 + dy_2^2 + dy_3^2 + \delta_{ij} dx^i dx^j \Bigr\},  \nonumber\\
\hat A_{[3]} &=& F_1^{-1} (dt - \omega_idx^i) \wedge dz_1 \wedge dz_2
   + U dy_1 \wedge dy_2 \wedge dy_3 \nonumber\\
  \!&+&\! F_2^{-1} (dt - \omega_i dx^i) \wedge dw_1 \wedge dw_2,
\eea
where  $U(x), F_2(x)$, are the harmonic functions on the $x^i$ space
while $\omega_i(x)$  and $F_1(x,w)$ satisfy the equations
\be
\ep_{ijk} \partial_{j} \omega_k = \partial_{i} U, \quad
\partial^2_{\vec x} F_1 + F_2 \partial^2_{\vec w} F_1 = 0.
\ee
This solution (contrary to its version without NUT) involves
the Chern-Simons term of (\ref{11d}). Direct check reveals the existence
of the $8D$ covariantly constant spinors corresponding to the
$1/4$ unbroken supersymmetry.

Thus we have given a purely vacuum interpretation to a certain class
of the $11D$ supergravity $M$-branes via new non-local duality between $11D$
supergravity and $8D$ vacuum Einstein gravity. It is worth noting that
the very natural correspondence between the Killing spinor equations
in both theories may have a deeper meaning in the context of $8D$
supergravities. In this connection it has to be emphasized that the type
of duality described here is by no means related to the direct dimensional
reduction from $11D$ to $8D$ theory.

One of the authors (DVG) would like to thank the Laboratory of Gravitation
and Cosmology of the University Paris-6 for hospitality and the CNRS
for support while this work was in progress.


\end{multicols}
\end{document}